\documentclass[journal]{IEEEtran}
\usepackage{amssymb,amsmath,epsfig,psfrag,epstopdf,enumerate}

\usepackage[english]{babel}
\usepackage{color}
\usepackage{graphicx}
\usepackage{epstopdf}

\usepackage[normalem]{ulem}
\usepackage{varioref}
\usepackage{balance}
\usepackage{subfig}

\usepackage{fancyhdr}
\fancypagestyle{firstpage}{
 \fancyhead[C]{N.B. A patent application has been filed in accordance with this work (and more inventive aspects). As a company, or investor, you are cordially invited to contact me at buygamnow@gmx.com (or pla@kth.se) for investment in, or acquisition of, the patent application. If you are aware of any potential investors, greatly appreciated if they are directed to this work.}
  \fancyfoot[C]{This work has been submitted to the IEEE for possible publication.  \\Copyright may be transferred without notice, after which this version may no longer be accessible.}
}

\newtheorem{definition}{Definition}[section]

\newtheorem{remark}{Remark}[section]
\newtheorem{theorem}{Theorem}[section]
\newtheorem{example}{Example}[section]

\begin{document}


\title{Golden Angle Modulation: \\ Geometric- and Probabilistic-shaping}


\author{
\IEEEauthorblockN{Peter Larsson \emph{Student
Member, IEEE.}}%
\thanks{The author is with the ACCESS Linnaeus Center and the School of
Electrical Engineering at  KTH Royal Institute of  Technology, SE-100 44
Stockholm, Sweden. (pla@kth.se)}
}
\maketitle
\thispagestyle{firstpage}

\begin{abstract}
Quadrature amplitude modulation (QAM), deployed in billions of communication devises, exhibits a shaping-loss of $\pi \mathrm{e}/6$ ($\approx 1.53$ dB) compared to the Shannon-Hartley theorem.  With inspiration gained from special (leaf, flower petal, and seed) packing arrangements (so called spiral phyllotaxis) found among plants, we have designed a shape-versatile, circular symmetric, modulation scheme, \textit{the Golden angle modulation (GAM)}. Geometric- and probabilistic-shaping-based GAM schemes are designed that practically overcome the shaping-loss of 1.53 dB. Specifically, we consider mutual information (MI)-optimized geometric-, probabilistic-, and joint geometric-and-probabilistic-GAM, under SNR-equality, and PAPR-inequality, constraints. Out of those, the joint scheme yields the highest MI-performance, and then comes the probabilistic schemes. This study finds that GAM could be an interesting candidate for future communication systems. Transmitter resource limited links, such as space probe-to-earth, satellite, and mobile-to-basestation, are scenarios where capacity achieving GAM could be of particular value.
\end{abstract}


\section{Introduction}
\IEEEPARstart{M}{odulation} schemes, in great number and variety, have been developed and analyzed in uncountable works. Examples of modulation formats are pulse amplitude modulation (PAM), square/rectangular quadrature amplitude modulation (QAM), phase shift keying (PSK), Star-QAM \cite{HanzoNgKellWebb04}, and amplitude-PSK (APSK) \cite{ThomasWeidDurr74}. Square-QAM, (or just QAM), is the de-facto-standard in existing wireless communication systems. However, at high signal-to-noise-ratio (SNR), QAM is known to asymptotically exhibit a 1.53 dB SNR-gap (a.k.a. shaping-loss)  between its mutual information (MI) performance and the additive white Gausian noise (AWGN) Shannon capacity (Shannon-Hartley theorem) \cite{ForneyUnger98}. This is attributed to the square shape and the uniform discrete distribution of the QAM-signal constellation points. APSK, inherently circular-symmetric, like a complex Gaussian random variable (r.v.), is of interest, but does not fully address the shaping-loss as such. Instead, geometric- and probabilistic-shaping techniques have been proposed to mitigate the shaping-gap \cite{ForneyUnger98}. An early work on geometric-shaping is nonuniform-QAM in \cite{BettsCaldeLaroi94}. Correspondingly, the papers on trellis shaping, \cite{Forney92}, and shell-mapping, \cite{KhandaniKaba93}, are early works on probabilistic-shaping. More recent works in this direction are, e.g., \cite{SzczecinskiAissGonzBaci06, MheichDuhaSzczMore11, XiangVale13, BuchaliIdleSchmaHu17}. Existing modulation formats and shaping methods have, in our view, not completely solved the shaping-loss issue, nor offered a modulation format that is practically suitable for this purpose. This motivates new modulation format(s) to be developed. 
\begin{figure}[tp!]
 \centering
 \vspace{-.4 cm}
  \includegraphics[width=9cm]{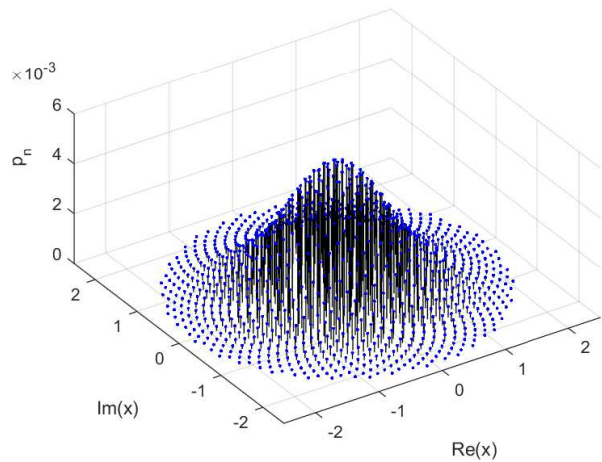}
 \caption{Probabilistic-bell-GAM signal constellation: \\ Minimum SNR - entropy-constrained formulation, $N=2^{10}$.}
 \label{fig:Fig5p5dot3}
 \vspace{-0.0cm}
\end{figure}
\begin{figure}[tp!]
 \centering
 \vspace{-.4 cm}
 \includegraphics[width=9cm]{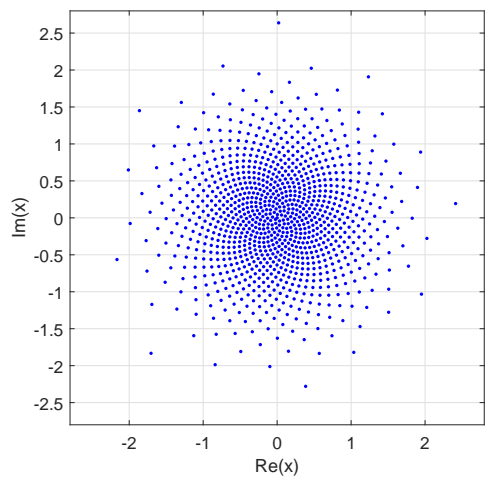}
 \caption{Geometric-bell-GAM signal constellation: \\ High-rate formulation, $N=2^{10}$.}
 \label{fig:Fig5p5dot2}
 \vspace{-0.35cm}
\end{figure}
In particular, we seek a modulation scheme with near (or essentially identical) AWGN Shannon capacity performance from the low SNR-range up to the SNR range where the MI approaches the entropy of the signal constellation. It is desirable that the proposed modulation scheme lends itself to both geometric- and probabilistic-shaping, to corresponding performance optimization, and to incorporate additional modulation design constraints (such as PAPR inequality constraints). Since probabilistic-shaping suggests additional, and generally more complex, hardware, compared to geometric-shaping, an objective of this work is to assess performance differences between the geometric- and the probabilistic-GAM approaches.

Inspired by the beautiful, and equally captivating, cylindrical-symmetric packing of scales on a cycad cone, the spherical-symmetric packing of seeds on a thistle seed head, or the circular-symmetric packing of sunflower seeds, we have recognized that this shape-versatile spiral-phyllotaxis packing principle, as observed among plants, is applicable to modulation signal constellation design.  In \cite{Larsson17a}, based on the observed spiral phyllotaxis packing principle, we introduced the idea of \textit{Golden angle modulation} (GAM). MI-performance of disc-shaped, and two geometrically-shaped, GAM schemes were treated. In the present work, we introduce and study three new probabilistic-bell-GAM (PB-GAM), and three new joint geometric-probabilistic-GAM (GPB-GAM), schemes. We also recapitulate the core idea(s) of GAM, and expand the study on geometric-shaping.
We describe, but also provide additional performance analysis, results, and extensions to, the disc-GAM and the GB-GAM schemes proposed in \cite{Larsson17a}. This offers a unified comprehensive treatment of GAM, setting the stage for comparison between geometric- and probabilistic-shaping approaches. Specifically, compared to \cite{Larsson17a}, disc-GAM is generalized to reduce its peak-to-average-ratio (PAPR) further, extended with symbol error rate (SER) analysis, and the idea of a constant magnitude code is discussed. The high-rate version of GB-GAM is extended with a SER analysis, and additional performance results are given. A new optimization formulation G2 of GB-GAM is introduced, which handles the optimization complexity issue with large constellations for optimization formulation G1. In the appendix, more detailed treatments of the optimization issues, not discussed in \cite{Larsson17a}, are also included.  Many promising extensions of GAM is foreseen, and we address some of those possibilities in this work.

The overall conclusion, noted in \cite{Larsson17a}, remains. Provided that the MI is less than the signal constellation entropy, and the constellation size increases, the MI-performance of GAM approaches the AWGN Shannon capacity. In this work, we see that this holds for both geometric-GAM and probabilistic-GAM. Optimization frameworks, developed here, shows that the MI-performance can be improved relative to results given in \cite{Larsson17a}. Our study suggests that probabilistic-shaped GAM offers slightly higher MI-performance than geometrically-shaped GAM. Joint optimization of both shaping-geometry and -probabilities is found to offer the greatest performance.

The paper is organized as follows. In Section \ref{GAM}, we first review the core-GAM signal constellation design. A PAPR-reducing generalized disc-GAM scheme is given in Section \ref{GDisc-GAM}. Geometric-shaping schemes are considered in Section \ref{sec:Sec5p5d2d2}. In Section \ref{sec:Sec5p5d2d3}, we propose and discuss GAM with probabilistic shaping. Joint geometric-probabilistic approaches are addressed in Section
\ref{sec:Sec5p5d2d4}. Numerical results are given in Section \ref{sec:Sec5p5d4}. In Section \ref{Summary}, we summarize and conclude this work.

\section{Golden Angle Modulation}
\label{GAM}
We first present the core design of GAM. This essentially follows \cite{Larsson17a}, where we first introduced GAM.

The core design of GAM builds on the use of the golden angle (or golden ratio) for phase rotations of consecutive constellation points. We define core-GAM as follows:
\begin{definition} (Golden angle modulation)
\label{def:Def5p5d1}
The probability of using the $n$th constellation point is denoted by $p_n$, and the complex amplitudes are
\begin{align}
x_n&=r_n\mathrm{e}^{i 2\pi \varphi n}, \, n\in\{1,2,\ldots,N\},
\end{align}
where $r_n$ is the radius of constellation point $n$, $2\pi\varphi$ denotes the golden angle in rads, and $\varphi=(3-\sqrt{5})/{2}$.
\end{definition}
We will assume that $r_{n+1}>r_n$ for an increasing spiral winding. For the probability, it may be equiprobable, $p_n=1/N$, or dependent on index $n$.
Hence, a constellation point, is located (the irrational number) $\varphi\approx0.382$ turns (or $137.5$ degrees) relative to the previous constellation point. Replacing $\varphi$, with $(1+\sqrt{5})/2\approx 1.618$, the golden ratio, gives an equivalent spiral winding, but in the opposite direction. More generally, $\varphi$ could of course be replaced with $k\pm(1-\sqrt{5})/2$, $k\in \mathbb{Z}$. Note that phase rotation value deviating with just $\approx1$\% from the golden angle (or ratio) destroys the, relatively, dense uniform packing.
The mathematical design of the phase rotation in Def. \ref{def:Def5p5d1} is inspired from the work by Vogel \cite{Vogel79}, who described an idealized growth pattern for the sunflower seeds, which in our notation is $x_n=\sqrt{n}\mathrm{e}^{i 2 \pi \varphi n}$. Vogel did however not consider, or see the parallel to, modulation design. More importantly, a key observation, and a central insight of our work, is to not restrict the radial function $r_n$ to $\sqrt{n}$, as in \cite{Vogel79}. This is what allow us to approximate, e.g., a complex Gaussian r.v. with geometric-shaping. I.e., tuning $r_n$, offers the dimension of geometric-shaping, whereas tuning $p_n$, gives the freedom of probabilistic-shaping.

GAM features the following advantages:
\begin{itemize}
\item
Natural constellation point indexing: In contrast to QAM, APSK, and other, without any natural index order, GAM enables a unique indexing based on signal phase, $2\pi \varphi n$, or magnitude, $r_n$, alone.
\item
Near-circular design: A circular design can offer enhanced MI-, distance-, SER- and PAPR- performance over a square-QAM design.
\item
Radial shape-flexibility: The radial distribution of constellation points can be tuned, while retaining an evenly distributed packing. We recognize this as a central feature of GAM which allows approximation of (practically) any circular-symmetric pdfs.
\item
Natural circular-symmetric probabilistic-shaping: The constellation design inherently lends itself for circular-symmetric probabilistic-shaping. We recognize this as a central feature of GAM which allows approximation of (practically) any circular-symmetric pdfs.
\item
Constellation alphabet size flexibility: Any number of constellation points can be used, while retaining the overall circular shape. This gives flexibility, e.g., in alphabet size of a channel coder, or a probabilistic-shaper.
\item
Rotation (and gain) invariance: The signal constellation has a uniquely identifiable phase and gain. This could, e.g., allow for blind channel estimation.
\end{itemize}

We also note that GAM has an average complex valued DC component. If this is considered a problem, the average DC component can be subtracted, or every second symbol can be negated, on average canceling the DC component. It is well-known that hexagonal packing is the densest 2-dimensional packing. This is desirable for the SNR-range when the MI is close to the signal constellation entropy, but otherwise of less interest. However, hexagonal packing does not offer the same features as listed above for GAM. Note that the index range is not absolute, but could, depending on convenience for the signal constellation design description, have a lower limit $\neq 1$.

\section{Generalized disc-GAM}
\label{GDisc-GAM}
Disc-GAM, as given in \cite{Larsson17a}, has a uniform disc-shaped distribution of constellation points. In some applications, the peak-to-average-ratio is of interest. Here, we generalize the disc-GAM design in \cite{Larsson17a}, thereby allowing the PAPR, on the expense of the MI-performance, to be controlled between 0 to 3 dB. We define the revised disc-GAM as follows:
\begin{definition} (Generalized disc-GAM)
\label{def:Def5p5d2}
The disc-GAM format, with average power $\bar P$, is characterized by
\begin{align}
r_n&=c_\textrm{disc}\sqrt{n}, \, n\in\{N_\textrm{l},N_\textrm{l}+1,\ldots,N_\textrm{h}\},\\
p_n&=\frac{1}{N}, \, \text{where}\\
c_\textrm{disc}&\triangleq \sqrt{\frac{2\bar PN}{N_\textrm{h}(N_\textrm{h}+1)-N_\textrm{l}(N_\textrm{l}-1)}},\\
N&\triangleq N_\textrm{h}-N_\textrm{l}+1.
\end{align}
\end{definition}
\begin{IEEEproof}
The proof of $c_\textrm{disc}$ is given in Appendix \ref{app:App5p5dA1}.
\end{IEEEproof}

We show the generalized disc-GAM constellation in Fig. \ref{fig:Fig5p5dot1}.

A few remarks about the generalized disc-GAM.
The entropy is simply $H_\textrm{disc}= \log_2 N$.
The PAPR is $PAPR_\textrm{disc}=2\bar P N N_\textrm{h} /\left(N_\textrm{h}(N_\textrm{h}+1)-N_\textrm{l}(N_\textrm{l}-1)\right)$. When $N_\textrm{l}=1$, we have regular disc-GAM. Then, when $N_\textrm{h}\rightarrow \infty$, $PAPR\simeq2$ $(\simeq 3)$ dB. If both $N_\textrm{h}$ and $N_\textrm{l}$, are large, the constellation concentrate on the rim of the disc. Thus, the PAPR can be controlled between 0 and 3 dB. This makes disc-GAM favorable over QAM which has an asymptotic $PAPR_\textrm{QAM}=4.8$ dB. This also makes generalized disc-GAM an interesting candidate to PSK, which has ${PAPR}_\textrm{PSK}=0$ dB, but with very poor MI-performance for large $N$. 
Focusing now on regular disc-GAM, with $N_\textrm{l}=1$. Letting the number of constellation points $N$ grows towards infinity, it is known that QAM asymptotically requires $10\log_2(\pi/3)$ ($\approx 0.2$ dB) higher average power than a uniform disc constellation (such as disc-QAM) for the same average distances between (uniformly packed) constellation points.
Also when number of constellation points $N$ grow towards infinity, it is found that QAM asymptotically requires $10\log_2(\pi/2)$ ($\approx 1.96$ dB) higher peak power than a disc (such as disc-QAM) for the same average distances between (uniformly packed) constellation points.

\subsubsection{Symbol Error Rate of disc-GAM}
While the focus in this work is on MI as performance measure, we briefly consider approximations to the symbol error rate (SER). Such approximation can be made in many ways. First, we note that the decision regions shape (and to some extent size) varies. As a first approximation, we assume that the decision regions are square-shaped. For large $N$, the majority of constellation points are not border-points, so border-effects can be neglected. The probability for that the $n$th point is in error is then $SER_n=1-\left(1-2Q\left(\sqrt{d_n^2/2\sigma^2}\right)\right)^2$, where $d_n^2$ is the decision region area. The total symbol error rate  is $SER=\frac{1}{N}\sum_{n=1}^N SER_n$. The area $d_n^2$ is estimated from the constellation area increase from point $n$ to point $n+1$, i.e $d_n^2\approx \pi r_{n+1}^2-\pi r_n^2$. With $r_n=c_\textrm{disc}\sqrt{n}$, we get 
\begin{align}
SER_\textrm{disc}&=1-\left(1-2Q\left( \sqrt{\frac{\pi S}{N+1}}\right)\right)^2.
\end{align}

\subsubsection{Disc-GAM Constant Magnitude Code}
In the above, modulation symbols have been treated as individual entities. Apart from employing channel coding to GAM, simpler codes can be constructed from multiple symbols. For example, we have realized that regular disc-GAM could also be used to construct constant  magnitude vector codes. For example, let a pair of disc-GAM-symbols form a two-symbol code, $\mathbf{x}_{n}=[x_{n}\ x_{n'}]$, where $n+n'=1+N, \, n\in\{1,2,\ldots,N\}$. Since $|\mathbf{x}_n|^2=
|c_\textrm{disc}|^2
|\begin{bmatrix}
\sqrt{n}\mathrm{e}^{2\pi \varphi n}\
\sqrt{N+1-n}\mathrm{e}^{2\pi \varphi (N+1-n)}
\end{bmatrix}|^2=2\bar P$, the sum-power is a constant, independent on index $n$.
The MI-performance for the constant modulus two-symbol code is far better than for (the one-symbol constant modulus) PSK, albeit not as good as the AWGN Shannon capacity. The Peak-to-average SNR Ratio is exactly 3 dB, compared to 0 dB for PSK. Generalization to higher dimensions is also possible. The reason why this design turn out to be appealing, is that in disc-GAM, the amplitude is $r_n\propto\sqrt{n}$. This implies that the squared magnitude, over any number of symbols, is always an integer.

\begin{figure}[tp!]
 \centering
 \vspace{-.4cm}
 \includegraphics[width=9cm]{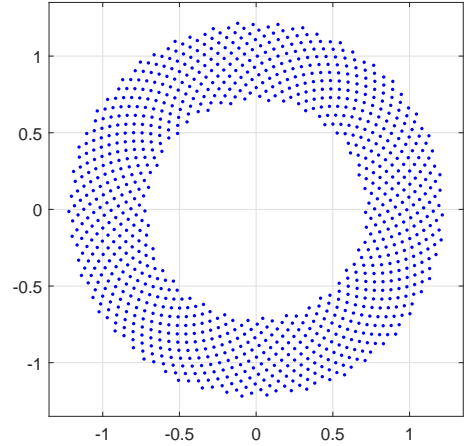}
 \caption{Generalized disc-GAM signal constellation, $N_\textrm{l}=2^{9}$ and $N_\textrm{h}=N_\textrm{l}-1+2^{10}$. $PAPR\approx 1.76$ dB, $H=10$ [b/Hz/s].}
 \label{fig:Fig5p5dot1}
 \vspace{0.0cm}
\end{figure}

\section{Geometric Bell-GAM}
\label{sec:Sec5p5d2d2}
In this section, we explore GAM with geometric-shaping.

\subsection{High-rate Approach}
\label{sec:Sec5p5d2d2d1}
In this, high-rate (HR), design, the aim is to approximate a continuous complex Gaussian pdf. The design can be derived from the inverse sampling theorem, and is given below.
\begin{theorem} (Geometric-bell-GAM high-rate)
\label{thm:Thm5p5d1}
Let $N$ be the number of constellation points, and $\bar P$ be the average power constraint. Then,  the complex amplitude of $n$th constellation point for the geometric-bell-GAM (GB-GAM)  is
\begin{align}
r_n&=c_\textrm{gb}\sqrt{\ln{\left(\frac{N}{N-n}\right)}}, \,n\in\{0,1,\ldots,N-1\}\\
p_n&=\frac{1}{N}, \, \text{where}\\
c_\textrm{gb}&\triangleq\sqrt{\frac{N \bar P}{N\ln{N}-\ln(N!)}}.
\label{eq:EqGaussrn}
\end{align}
\end{theorem}
\begin{IEEEproof}
The proof is given in Appendix \ref{app:App5p5dA2d1}.
\end{IEEEproof}

We illustrate the GB-GAM signal constellation in Fig. \ref{fig:Fig5p5dot2}, and note that it is densest at its center, i.e. where the pdf for the complex Gaussian r.v. peaks.

Some remarks about the HR-formulation of GB-GAM. When $N \rightarrow \infty$, since $\lim_{N\rightarrow \infty} \frac{1}{N} \ln{\left(\frac{N^N}{N!}\right)}=1$, asymptotically, we get $r_n\simeq \sqrt{\bar P \ln{\left(\frac{N}{N-n}\right)}}$.
The entropy is $H_\textrm{gb}= \log_2 N$.
The PAPR is $PAPR_\textrm{gb}=c_\textrm{gb}^2\ln(N)
=\bar P /(1-\ln{(N!)}/N \ln {N})\simeq \bar P \ln{(N)}$, which tends to infinity with $N$. This is expected as the PAPR of a Gaussian distributed r.v. is infinite.
Note that the index range in \eqref{eq:EqGaussrn} need to be $n\in\{0,1,\ldots,N-1\}$, since $r_N=\infty$.

\subsubsection{Symbol Error Rate}
A similar SER-analysis, as for disc-GAM, involves the same relations, but the areas $d_n^2$ of the decision regions differ. The area for each constellation point $n$ can be approximated by the differential area increase $d_n^2\approx \pi (r_{n+1}^2- r_n^2)= \pi c_\textrm{gb}^2\ln\left(\frac{N-n}{N-(n+1)}\right)$. The SER can then be approximated as
\begin{align}
SER_\textrm{gb}&=\frac{4}{N}\sum_{n=0}^{N-1}Q\left( f_\textrm{gb}\right)-Q\left(f_\textrm{gb}\right)^2,\\
f_\textrm{gb}(n,N,S)&\triangleq \sqrt{\frac{SN\pi \ln\left(\frac{N-n}{N-(n+1)}\right)}{2(N\ln(N)-\ln(N!))}}.
\end{align}
For the SER, $1-(1-2Q(x))^2=4Q(x)-4Q(x)^2$ was used.

\subsection{MI-optimization of GB-GAM: Formulation-G1}
\label{sec:Sec5p5d2d3d1}
While MI-performance is the prime performance of interest, the GB-GAM high-rate approach does not specifically aim to maximize the MI (but rather approximate a complex Gaussian r.v.). Moreover, $N$ is not infinite, as for the high-rate assumption, but limited in practice. We also note that the magnitudes of $r_N$ for the high-rate approach appears "unnecessarily large". For those reasons, we now explore an MI-optimization formulation for GAM.

In this method, we let $p_n=1/N$, and vary $r_n$ in order to maximize the MI, for a desired SNR, $S$. The optimized signal constellation points are $x_n^*=r_n^* \mathrm{e}^{i 2\pi \varphi n}$. More formally, allowing for a complex valued output r.v. $Y$, and a complex valued (discrete modulation) input r.v. $X$, the optimization problem is
\begin{equation}
\begin{aligned}
& \underset{r_n}{\text{maximize}}
& & I(Y;X), \\
& \text{subject to}
& & r_{n+1} \geq r_{n}, \; n = \{1,2, \ldots, N\},\\
& & & r_1 \geq 0,\\
& & & \sum_{n=1}^{N}\frac{p_nr_n^2}{\sigma^2}=S.
\end{aligned}
\end{equation}

\begin{remark}
\label{rm:Rm5p5d9}
In some cases, a PAPR-criteria is of interest. For this reason, we simply add a PAPR inequality constraint, $PAPR\leq PAPR_0$, to the above optimization problem, where $PAPR_0$ is the target PAPR. This can be expressed as
\begin{align}
\left(\frac{1}{PAPR_0}-p_N\right)r_{N}^2-\sum_{n=1}^{N-1}p_nr_n^2\leq 0.
\end{align}
When the PAPR inequality constraints is included, by the constellation design thin out in the center, and concentrate as a ring, somewhat similar to Fig. \ref{fig:Fig5p5dot1} for generalized disc-GAM.
Note that a PAPR inequality constraint, as discussed, could be included in any given GAM-optimization formulation.
\end{remark}

In the above, and in the following, the MI can be expressed in terms of the differential entropies $h(Y)$, and conditional differential entropy $h(Y|X)$, as $I(Y;X)=h(Y)-h(Y|X)=h(Y)-h(W)$, where $h(W)=\log_2(\pi \mathrm{e}\sigma^2)$, $h(Y)=-\int_{\mathbb{C}} f_Y \log_2(f_Y) \, \mathrm{d}y$, integrating over the complex domain, with $f_Y=\sum_{n=1}^{N} f(y|x_n)p_n=\frac{1}{\pi\sigma^2}  \sum_{n=1}^{N} p_n \mathrm{e}^{-\frac{|y-x_n|^2}{\sigma^2}}$.

The optimization problem G1 is hard to solve analytically. In Appendix \ref{app:App5p5dA2d1}, we (partially) illustrate why it is hard to solve the problem. In Section \ref{sec:Sec5p5d4}, due to complexity, we resort to a numerical optimization solver for $N=16$ to illustrate the improvement over the HR-formulation.

\subsection{MI-optimization of GB-GAM: Formulation-G2}
\label{sec:Sec5p5d2d3d2}
A practical issue with the optimization formulation-G1 is that $N$ undetermined $r_n$ need to be solved for. This makes numerical optimization computationally intensive, and time consuming. An alternative approach is to let a function describes a continuous outward-growing spiral, and tune just a few parameters to maximize the MI. Thereby, the complexity and optimization time are reduced. This approach, which is more constrained than formulation-G1, should, if properly designed, produce a negligibly reduced optimal MI compared to formulation-G1.
Therefore, we consider a positive increasing (spiral power) function $f_\textrm{P}(x,c_0,c_1\ldots,c_K), \, x\in(0,1)$, and let the magnitude for the $n$th signal point be $r_n=\sqrt{f_\textrm{P}(n/N)}$.\footnote{The motivation for the square root of a spiral power function stems from disc-GAM-, and GB-GAM (HR). For small $n$, large $N$, the GB-GAM (HR) has the same form as disc-GAM, since $\sqrt{-\ln(1-n/N)}\simeq\sqrt{n/N}$.} We also want to ensure that $f_\textrm{P}(x)$ is positive and growing. This is e.g. satisfied with $f_\textrm{P}(0)\geq 0$ and $f'_\textrm{P}(x)> 0$, where $f'_\textrm{P}(x)\triangleq \mathrm{d}f_\textrm{P}(x)/\mathrm{d}x$. With this, and $p_n=1/N$, the optimization problem is then
\begin{equation}
\begin{aligned}
& \underset{c_k}{\text{maximize}}
& & I(Y;X), \\
& \text{subject to}
& & f'_\textrm{P}(n/N,c_k) \geq 0, \;  n = \{1,2, \ldots, N\},\\
& & & f_\textrm{P}(1/N,c_k)\geq 0, \;  k= \{1,2, \ldots, K\},\\
& & & \sum_{n=1}^{N}\frac{f_\textrm{P}(n/N,c_k)}{N\sigma^2}=S.
\end{aligned}
\end{equation}
The optimal $n$th point magnitude is then $r_n^*=\sqrt{f_\textrm{P}\left(n/N,c_0^*,c_1^*,\ldots,c_K^*\right)}$. The positive derivative criteria (for spiral growth) can alternatively have been formulated as $f_\textrm{P}((n+1)/N,c_k) \geq f_\textrm{P}(n/N,c_k), \;  n = \{1,2, \ldots, N\}$.

\begin{example}
\label{ex:Ex5p5d1}
As an example, we consider a $K$th degree polynomial spiral power function $ f_\textrm{P}(x)=\sum_{k=0}^K c_k x^k$.
The optimization problem, using the derivative-based constraint, and including a PAPR inequality constraint, is now
\begin{equation}
\begin{aligned}
& \underset{c_k}{\text{maximize}}
& & I(Y;X), \\
& \text{subject to}
& & \sum_{k=1}^K c_k k\left(\frac{n}{N}\right)^{k-1} > 0, \; n = \{1,2, \ldots, N\},\\
& & & \sum_{k=0}^K c_k \left(\frac{1}{N}\right)^k\geq 0,\\
& & & \sum_{n=1}^{N}\sum_{k=0}^K \frac{c_k}{N\sigma^2} \left(\frac{n}{N}\right)^k=S,\\
& & & \sum_{k=0}^K c_k-PAPR_0\sum_{n=1}^{N}\sum_{k=0}^K \frac{c_k}{N} \left(\frac{n}{N}\right)^k\leq 0.
\end{aligned}
\end{equation}
Here, we have intentionally designed linear constraints in $c_k$, but non-linear in the objective function. This eases implementation with an optimization solver, e.g. \textrm{MATLAB}'s \textit{fmincon}.
\end{example}
This problem, even with the complexity-reducing polynomial formulation, is also hard to solve analytically. Hence, in Section \ref{sec:Sec5p5d4}, we resort to a numerical optimization solver. However, for sufficiently large $N$, the complexity with the polynomial approach is generally much lower than for optimization formulation-G1. In Appendix \ref{app:App5p5dA2d2}, we illustrate the analytical challenges with this optimization problem. 
Note that other functions, than a polynomial, could be used. A Pade' approximant, such as
$f_\textrm{P}(x)=(a_1x+a_2x^2)/(1+b_1x)$, could be an option.

\section{Probabilistic Bell-GAM}
\label{sec:Sec5p5d2d3}
We now turn our attention to probabilistic-shaping. Here, we assume the disc-GAM signal constellation design with $r_n=c_\textrm{pb} \sqrt{n}$ and assign $p_n$ such that the pmf, in some sense, tend to approximate a bell-shaped complex Gaussian pdf. We denote such scheme probabilistically-shaped bell-GAM (PB-GAM). We analytically develop an PB-GAM signal constellation which is SNR-independent, and also formulate MI-optimized SNR-dependent variants.

\subsection{Minimum SNR with Entropy-constraint}
\label{sec:Sec5p5d2d4d1}

Our first take on this is to optimize $p_n$  such that an entropy constraint, $H_\textrm{pbse}$, is fulfilled, while minimizing the SNR. The optimization problem is
\begin{equation}
\begin{aligned}
& \underset{p_n}{\text{minimize}}
& & \sum_{n=1}^{N}\frac{p_nr_n^2}{\sigma^2},  \;  n = \{1,2, \ldots, N\}, \\
& \text{subject to}
& & -\sum_{n=1}^N p_n \log_2 {p_n} = H_\textrm{pbse},\\
& & & \sum_{n=1}^N p_n = 1,
\end{aligned}
\end{equation}
where we let $r_n=c_\textrm{pb} \sqrt{n}$.
Using Lagrangian optimization, the optimal solution is given by the following theorem.
\begin{theorem}
(Probabilistic-bell-GAM with minimum SNR and an entropy constraint)
\label{thm:Thm5p5d2}
Let $N$ be the number of constellation points, $p_n$ is the probability that constellation point $n\in\{1,2,\ldots,N\}$ is used,  $H_\textrm{pbse}$ is the entropy constraint, and $\bar P$ is average power. Then, PG-GAM is characterized by
\begin{align}
r_n&=c_\textrm{pbse}\sqrt{n} , \, n\in\{1,2,\ldots,N\},\\
c_\textrm{pbse}&\triangleq \sqrt{\bar P \left(\frac{1}{1-\xi}-\frac{N\xi^N}{1-\xi^N}\right)^{-1}},\\
p_n&=\frac{1-\xi}{1-\xi^N} \xi^{n-1},\label{eq:Eqpnpg}\\
H_\textrm{pbse}&=-\ln \left( \frac{1-\xi}{1-\xi^N}\right)+\left(\frac{N\xi^N}{1-\xi^N}-\frac{\xi}{1-\xi}\right)\ln {(\xi)},
\label{eq:EqHpg}
\end{align}
where $\xi$ is a constant determined from \eqref{eq:EqHpg} given a desired entropy $H_\textrm{pbse}$ in nats.
\end{theorem}
\begin{IEEEproof}
The proof is given in Appendix \ref{app:App5p5dA3}.
\end{IEEEproof}

We see that when $N \rightarrow \infty$, the following asymptotic results holds;
$r_n\simeq\sqrt{\bar P (1-\xi)}\sqrt{n}$, $p_n\simeq(1-\xi)\xi^{n-1}$, and $H_\textrm{pbse}\simeq -\ln \left( 1-\xi \right)+\left(\frac{\xi}{1-\xi}\right)\ln {(\xi)}$. 
We remark that this scheme needs a shaping (or channel) encoder that outputs symbol indices with probabilities given by the geometric pmf \eqref{eq:Eqpnpg}. This may be a challenge, but we discuss possible alternative in Section \ref{sec:Sec5p5d2d4}. The design dictates that $H<\log_2(N)$, and practically we find that $H=\log_2(N)-1$ is a good choice.  A potential advantage of this scheme, in terms of a probabilistic shaper design, is that $p_n$, and $H$, are fixed wrt SNR for a given $N$.

In Fig. \ref{fig:Fig5p5dot3}, we illustrate the PB-GAM constellation  with probabilistic-shaping in form of a 3D-side-view. We observe the expected discretized bell-shape, with the larger discrete probability values at its center, where the pdf for a continuous complex Gaussian r.v. peaks.

While we have found an analytical, and SNR independent, optimized solution, we are again interested of maximizing the MI-performance that best exploits the given number of constellation points. This is the aim of the next scheme.

\subsection{MI-optimization of PB-GAM: Formulation-P1}
\label{sec:Sec5p5d2d5d1}
In this method we optimize $p_n$ which maximizes the MI for a given SNR. We assume that $r_n$ is given and based on the disc-GAM format $r_n=c_\textrm{pbis}\sqrt{n}$. The optimization problem is
\begin{equation}
\begin{aligned}
& \underset{p_n}{\text{maximize}}
& & I(Y;X), \\
& \text{subject to}
& & \sum_{n=1}^N p_n = 1, \; n = \{1,2, \ldots, N\},\\
& & & \sum_{n=1}^{N}\frac{p_nc_\textrm{pbis}^2 n}{\sigma^2}=S.
\end{aligned}
\end{equation}
While this optimization problem only involves equality constraints, and the Lagrangian and its derivative are easily given, it is still analytically untractable.\footnote{The intuitively-based condition, $p_{n+1} < p_{n}$, when approximating a complex Gaussian r.v., could also be added to the optimization problem.}

\subsection{MI-optimization of PB-GAM: Formulation-P2}
\label{sec:Sec5p5d2d5d1b}
In formulation-G2, we simplified the optimization of formulation-G1 by optimizing just a few variables. Here, we do the same for probabilistic-shaping.
For the disc-GAM constellation with $r_n=c_\textrm{d}\sqrt{n}$, the cell-sizes are approximately of the same size for a large $N$. Hence, the discrete probabilities are
\begin{align}
p_n
&\approx c_\textrm{d}\frac{1}{\pi \sigma^2}\mathrm{e}^{-\frac{r_n^2}{\sigma^2}} =c_\textrm{d}\frac{1}{\pi \sigma^2}\mathrm{e}^{-\frac{c_\textrm{d}^2n}{\sigma^2}}. \notag
\end{align}
Normalizing the sum-probability gives,
\begin{align}
p_n
&=\frac{1-\mathrm{e}^{-\frac{c_\textrm{d}^2}{\sigma^2}}} {1-\mathrm{e}^{-\frac{c_\textrm{d}^2}{\sigma^2}N}}
\mathrm{e}^{-\frac{c_\textrm{d}^2}{\sigma^2}(n-1)}
=\frac{1-\xi}{1-\xi^N}\xi^{n-1},
\end{align}
where $\xi=\mathrm{e}^{-\frac{c_\textrm{d}^2}{\sigma^2}}$, and the probability is identical to the probability in Theorem \ref{thm:Thm5p5d2}. If $\xi<<1$, $p_n$ has a characteristic bell-shape. On the other hand, $\lim_{\xi\rightarrow 1}p_n=1/N$, which implies a disc-shaped pmf. Hence, we have a pmf that can, depending on $\xi=(0,1)$, approximate and smoothly be adjusted between a truncated complex Gaussian-shaped pmf and a disc-shaped pmf. In the SNR-limited range, we intuitively desire a complex Gaussian like pmf approximation, whereas in the entropy limited range, we require a disc-shaped pmf, to maximize MI.
Hence, we can optimize the pmf wrt $\xi$ to maximize the MI. The optimization problem, with the SNR equality constraint determined in $\xi$, can then be written
\begin{equation}
\begin{aligned}
& \underset{\xi}{\text{maximize}}
& & I(Y;X), \\
& \text{subject to}
& & -\ln(\xi)\left(\frac{N \xi^N}{1-\xi^N}-\frac{1}{1-\xi}\right)=S.
\end{aligned}
\end{equation}
This scheme fully use all constellation points for the maximum entropy, i.e. asymptotically, for increasing SNR, ${H}\simeq\log_2{N}$. A potential disadvantage, similar to P1, in terms of designing a probabilistic shaper, is that $p_n$, depends on the SNR.

\section{Joint Probabilistic-Geometric Bell-GAM}
\label{sec:Sec5p5d2d4}
In the preceding sections, we have proposed to use either geometrically-shaped GAM, or probabilistically-shaped GAM. In the following, we avoid such restrictions, and consider hybrid-shaping schemes.

\subsection{Optimized Geometric-shaping with Given $p_n\neq 1/N$}

Building on GB-GAM with optimized magnitudes, such as the G1- and G2-formulation, we allow for a given pmf where $p_n\neq 1/N$. This corresponds to the case where we have a channel encoder, source coder, or shaper, with a given pmf that can not fully approximate a complex Gaussian r.v., but some inherent probabilistic-shaping gain is nevertheless provided. A geometric-shaping, optimizing the magnitudes of the signal constellation, can then fine-tune the MI-performance. 

Generally we expect that $p_{n+1}< p_n, \forall n$, but such shaper may be hard to design. Briefly noting that using a suboptimal Huffman coding based shaper, we can approximate the complex Gaussian pdf with a pmf in form of a circular (multi-level) "wedding-cake-shape" where $p_n$ are chosen to be powers of $1/2$, and then optimize the signal constellation magnitudes.

\subsection{Optimized Probabilistic-shaping with Given $r_n\neq c_\textrm{pb}\sqrt{n}$}

Similarly, $r_n\neq c_\textrm{pb}\sqrt{n}$ can be assumed is given, and the probabilities for a probabilistic-shaper are optimized instead. For this purpose, the P1- and P2-optimization formulations can be used with a given geometric distribution.

\subsection{MI-optimization of PGB-GAM: Formulation-GP1}
\label{sec:Sec5p5d2d6d1}
Finally, we propose a method involving joint geometric-probabilistic optimization. More precisely, $r_n$ and $p_n$ are optimized to maximize the MI under an SNR-constraint. The $n$th constellation point is $x_n=r_n\mathrm{e}^{i 2 \pi n}$ with probability $p_n$.
The optimization problem is then
\begin{equation}
\begin{aligned}
& \underset{p_n, r_n}{\text{maximize}}
& & I(Y;X), \\
& \text{subject to}
& & \sum_{n=1}^N p_n = 1, \; n = \{1,2, \ldots, N\},\\
& & & r_{n+1} > r_{n}, \\
& & & r_{1} \geq 0, \\
& & &  \sum_{n=1}^{N}\frac{p_n r_n^2}{ \sigma^2}= S.
\end{aligned}
\end{equation}
An aesthetic constraint of decreasing probabilities, $p_{n+1} < p_{n}$, could be added to the optimization problem. However, the MI-performance is, by necessity (negligibly) reduced.

Again, an analytical solution is untractable. In Section \ref{sec:Sec5p5d4}, we opt for numerical non-linear optimization with an optimization solver. We are able to handle up to $N=16$, thereby illustrating improved MI-performance over both geometric- and probabilistic-shaping only.

\section{Numerical Results and Discussions} 
\label{sec:Sec5p5d4}
In this section, the MI-performance of some of the presented GAM-schemes are studied.

\subsection{Mutual Information Performance}
\label{sec:Sec5p5d4d1}
To exactly analyze the MI of GAM (with its irregular cell-shapes and cell-sizes) is untractable. Except for the numerical optimized constellations, which uses numerical integration, we settle for Monte-Carlo simulation of the MI-performance curves. The MI estimator is
\begin{align}
\hat I(Y;X)&=\frac{1}{K_\textrm{mc}}
\sum_{\kappa=1}^{K_\textrm{mc}} \log_2\left( \frac{p\left(y^{(\kappa)}|x^{(\kappa)}\right)} {\sum_{n=1}^N p\left(y^{(\kappa)}|x_n\right)p(x_n)}\right),
\label{eq:MIest}
\end{align}
where $K_\textrm{mc}$ is the number of  iterations, using the baseband samples $y^{(\kappa)}=x^{(\kappa)}+w^{(\kappa)}$
and the bivariate conditional pdf is
$p(y|x)=\frac{1}{\pi \sigma^2}\mathrm{e}^{-\frac{|y-x|^2}{\sigma^2}}$.
For each iteration $\kappa$, $x$ is randomly selected from $x_n$ with probabilities $p(x_n)=p_n$. We let the average power of the modulation signal be unit-normalized, and the noise variance is $\sigma^2=1/S$.

\subsection{Geometric Bell-GAM}
\label{sec:Sec5p5d4d1d1}

\begin{figure}[tp!]
 \centering
 \vspace{-.4 cm}
 \includegraphics[width=9cm]{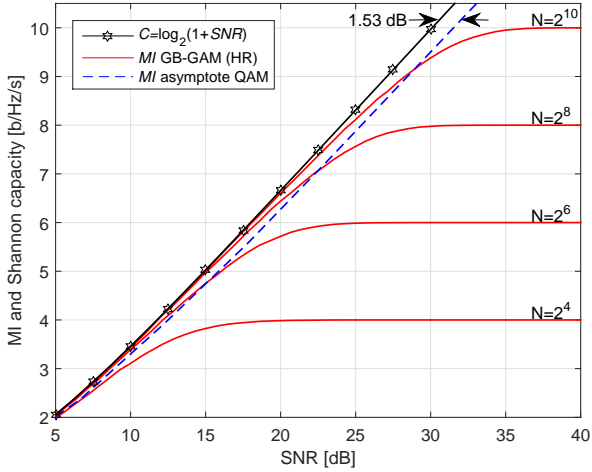}
 \caption{MI of GB-GAM (high-rate) with $\log_2(N)=\{4,6,8,10\}$, QAM asymptote, and AWGN Shannon capacity.}
 \label{fig:Fig5p5dot4}
 \vspace{0.0cm}
\end{figure}

In Fig. \ref{fig:Fig5p5dot4}, we illustrate the MI-performance for the high-rate GB-GAM version together with the AWGN Shannon capacity. As expected, for larger constellation size $N$, a greater overlap with the AWGN Shannon capacity is seen. The MI-performance matches the capacity curve for, say $MI \lessapprox H-2$. Naturally, the MI is limited by the entropy of the signal constellation. We observe an intermediate SNR region, where $MI\lessapprox \min{\left(\log_2(1+S),\log_2(N)\right)}$, and further constellation optimization is of interest.

\begin{table}[tp!]
\normalsize
\begin{center}
  \begin{tabular}{|r *{4}{|c}|}
    \hline
        \textbf{SNR}
    &
        \textbf{AWGN capacity}
    &
        \textbf{HR}
    &
        \textbf{G1}
    &
        \textbf{G2}
    \\
    \hline
    $\approx4.8$ dB&2&1.921&1.961&1.947\\
    \hline
    $\approx11.8$ dB&4&3.440&3.549&3.542\\
    \hline
    $15$ dB&$\approx 5.03$&3.828&3.926&3.921\\
    \hline
  \end{tabular}
\caption{MI in [b/Hz/s] for HR-, G1-, G2-formulations of geometric bell-GAM with $N=16$. For G2, $f_\textrm{P}$ is a third-order polynomial.}
\label{tab:Tab1d5p5d1}
\end{center}
\vspace{-0.35cm}
\end{table}

\begin{table}[tp!]
\normalsize
\begin{center}
  \begin{tabular}{|r *{3}{|c}|}
    \hline
        \textbf{SNR}
    &
        \textbf{AWGN capacity}
    &
        \textbf{HR}
    &
        \textbf{G2}
    \\
    \hline
    $\approx4.8$ dB&2&1.997&1.997\\
    \hline
    $\approx11.8$ dB&4&3.972&3.965\\
    \hline
    $\approx24.1$ dB&8&7.403&7.528\\
    \hline
    $33$ dB&$\approx 10.963$&7.999&8.000\\
    \hline
  \end{tabular}
\caption{MI in [b/Hz/s] for HR-,  G2-formulations of geometric bell-GAM with $N=256$. For G2, $f_\textrm{P}$ is a third-order polynomial.}
\label{tab:Tab1d5p5d2}
\end{center}
\vspace{-0.35cm}
\end{table}

In Tab. \ref{tab:Tab1d5p5d1}, the MI of GB-GAM with the HR-, G1-, and polynomial G2-formulations are given. As expected, the optimized schemes, G1 and G2, perform better than the HR-case. Moreover, G1, where each constellation point is optimized, offers higher MI than the G2-case with its fewer degrees of freedom for optimization.

In Tab. \ref{tab:Tab1d5p5d2}, we give the MI-performance for the HR-formulation, and the third order polynomial G2-formulation for a larger constellation size, $N=256$. The G1-formulation is not considered due to complexity, as $N$ is too large. The MI of the G2-formulation is generally higher than for the HR-formulation, except at low SNR where the polynomial spiral power function approximation is less suitable.

\begin{figure*}%
\centering
\hspace{-30pt}%
\subfloat[][]{%
\label{fig:MagnG2-b}%
 \includegraphics[width=6.5cm]{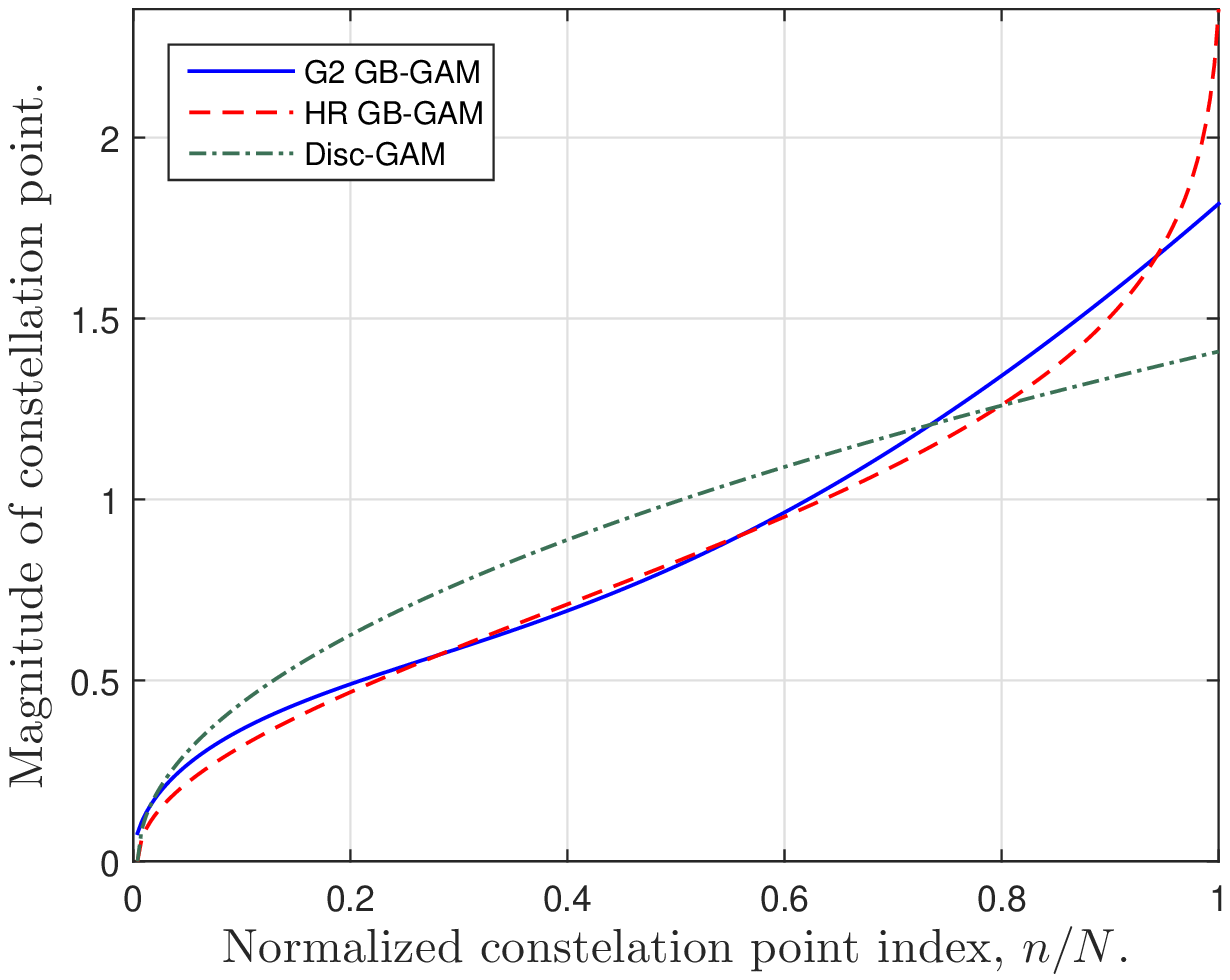}}
\hspace{-10pt}%
\subfloat[][]{%
\label{fig:MagnG2-c}%
 \includegraphics[width=6.5cm]{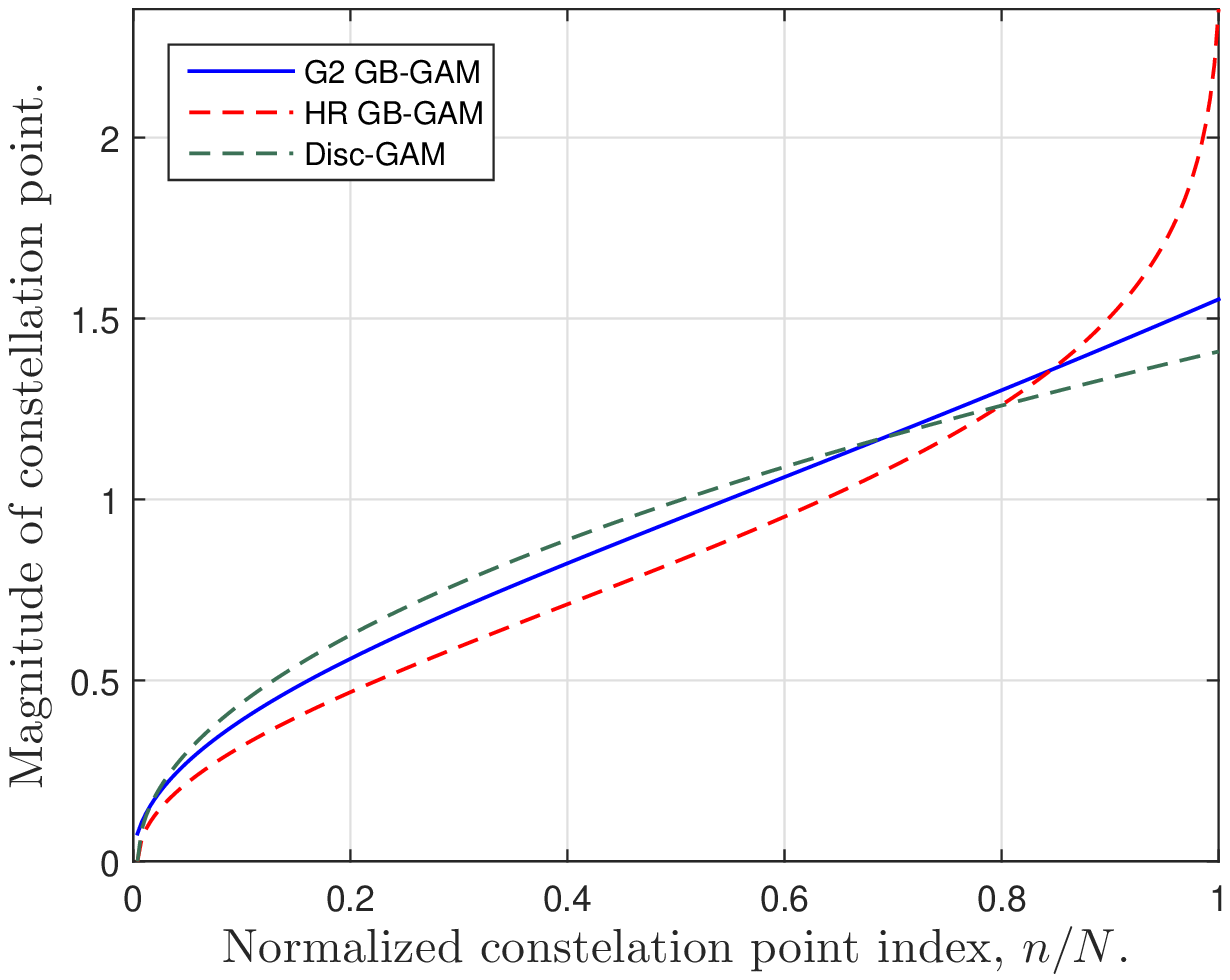}}%
\hspace{-10pt}%
\subfloat[][]{%
\label{fig:MagnG2-d}%
 \includegraphics[width=6.5cm]{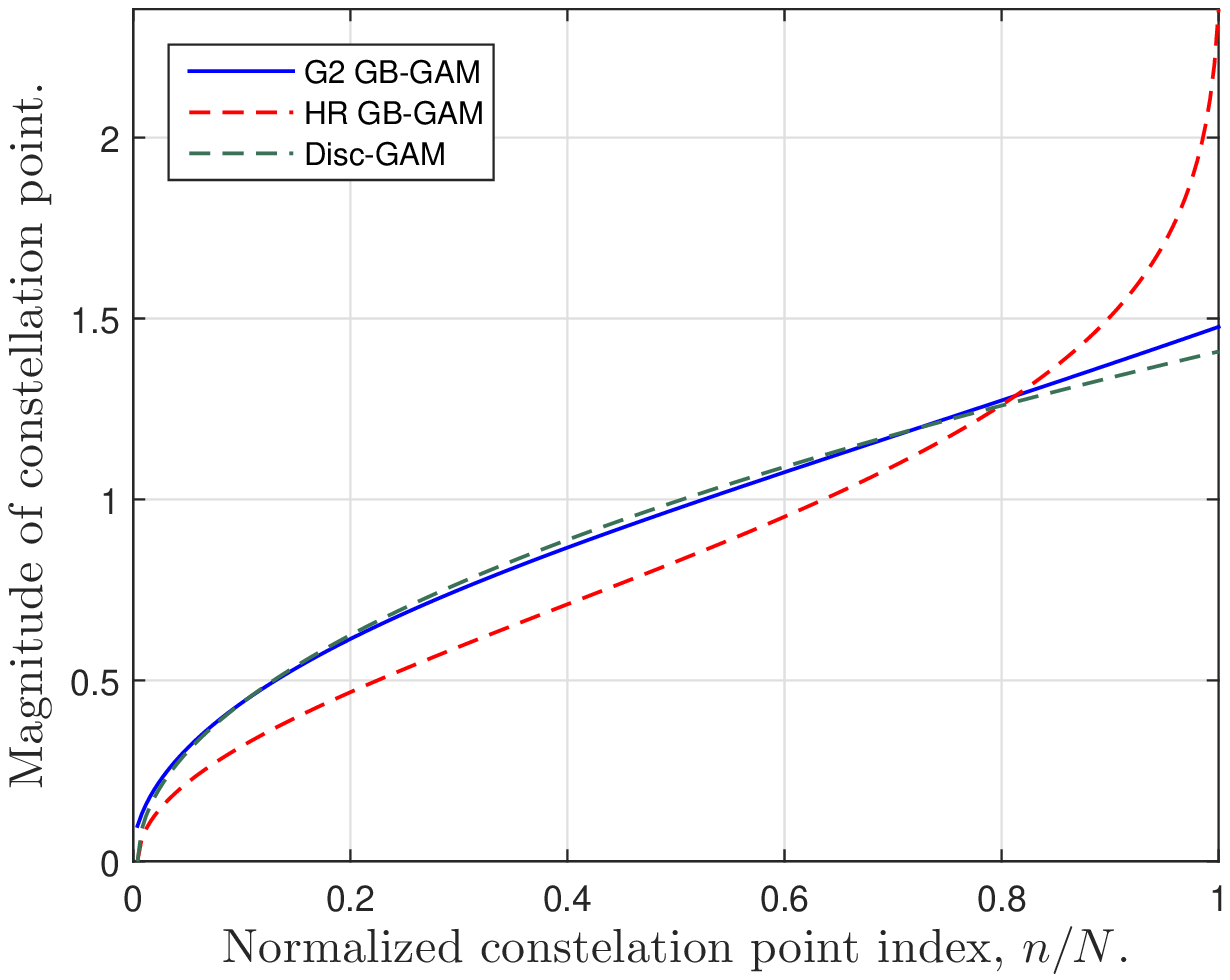}}%
\caption[ .]{Constellation point magnitude distribution of GB-GAM G2, with third-order polynomial spiral power function, and $N=256$:
\subref{fig:MagnG2-b} $S\approx 11.8$ dB;
\subref{fig:MagnG2-c} $S\approx 24.1$ dB; and,
\subref{fig:MagnG2-d} $S=33$ dB.}%
\label{fig:MagnG2}%
\end{figure*}

\begin{figure*}%
\centering
\hspace{-30pt}%
\subfloat[][]{%
\label{fig:ConstG2-b}%
 \includegraphics[width=7cm]{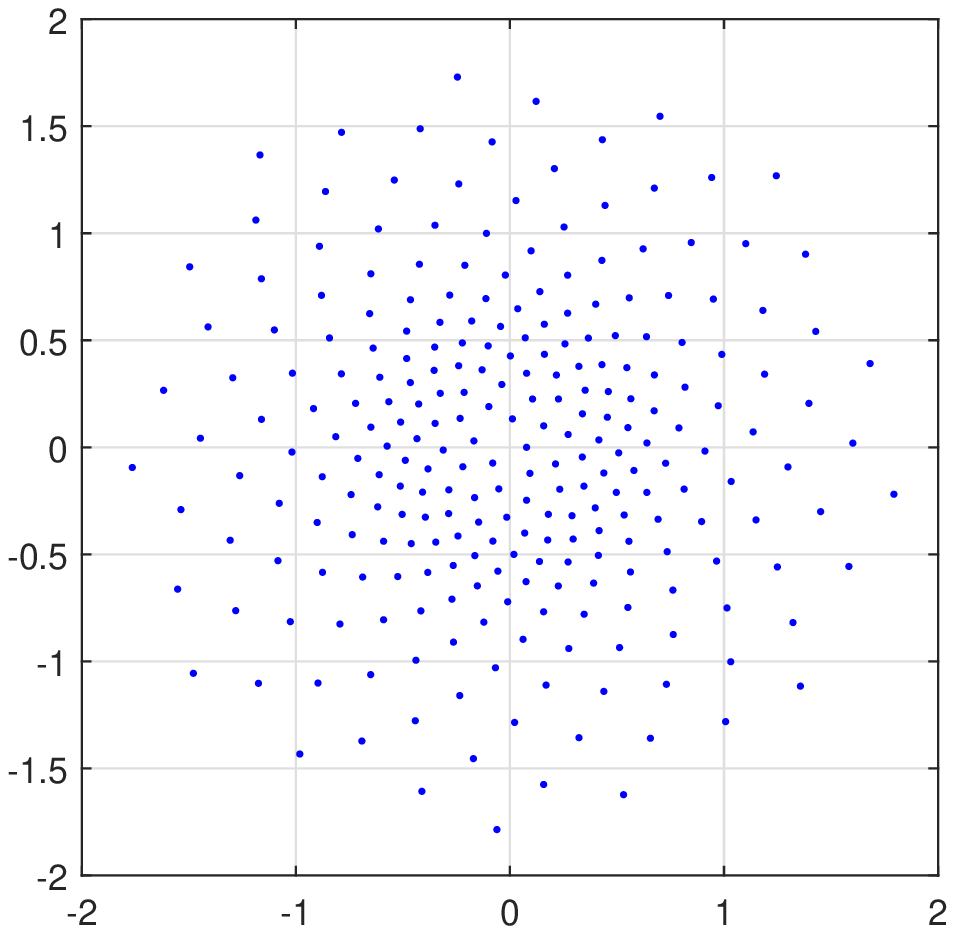}}
\hspace{-30pt}%
\subfloat[][]{%
\label{fig:ConstG2-c}%
 \includegraphics[width=7cm]{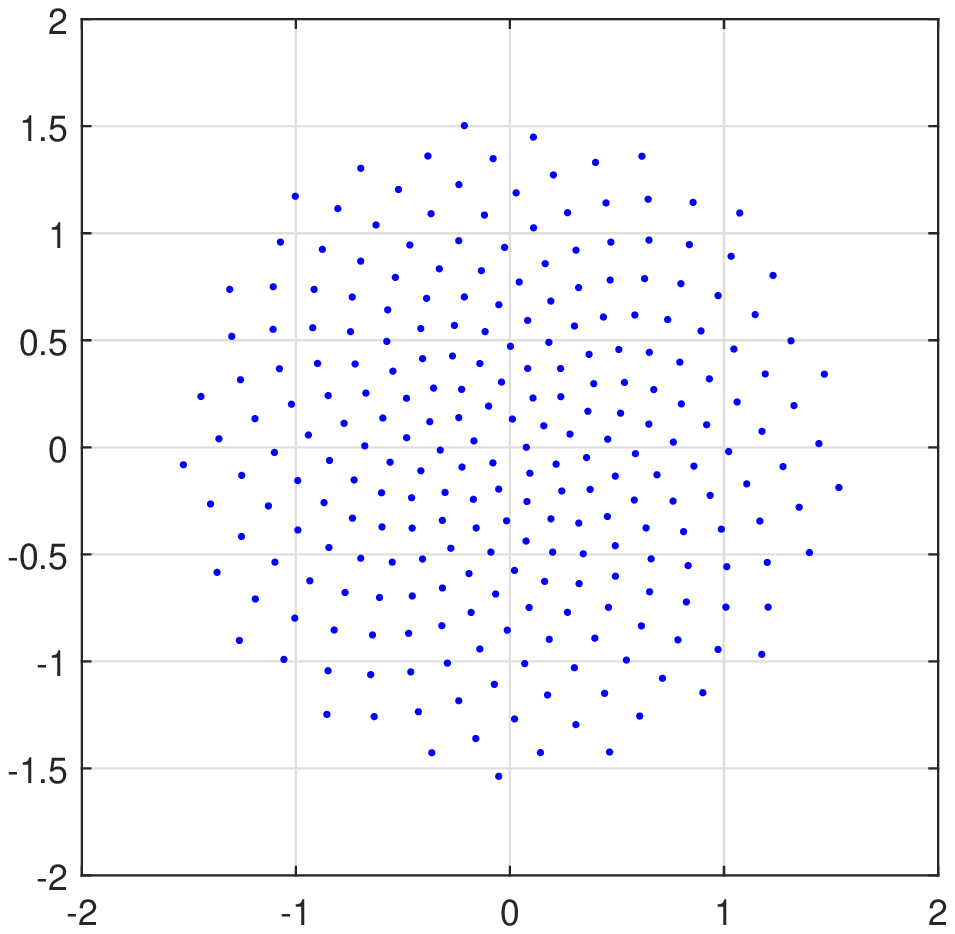}}%
\hspace{-30pt}%
\subfloat[][]{%
\label{fig:ConstG2-d}%
 \includegraphics[width=7cm]{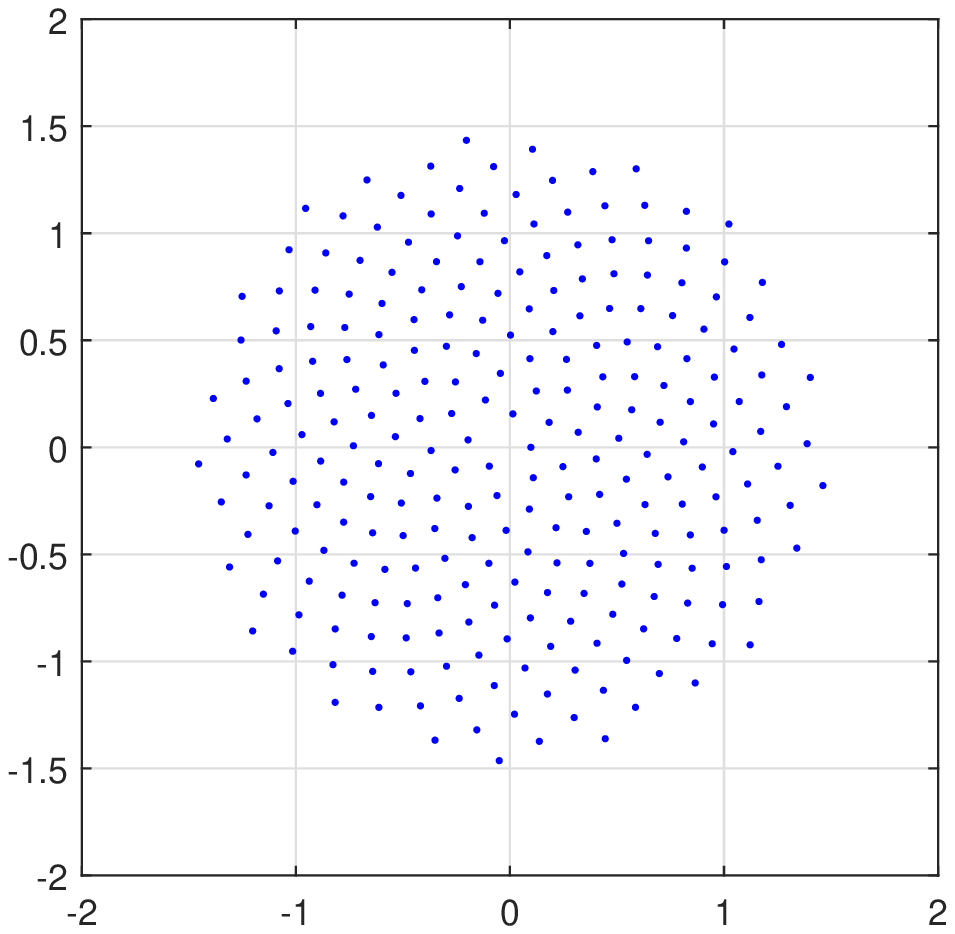}}%
\caption[ .]{Signal constellations of GB-GAM G2, with third-order polynomial spiral power function, and $N=256$:
\subref{fig:ConstG2-b} $S\approx 11.8$ dB;
\subref{fig:ConstG2-c} $S\approx 24.1$ dB; and,
\subref{fig:ConstG2-d} $S=33$ dB.}%
\label{fig:ConstG2}%
\end{figure*}

In Fig.~\ref{fig:MagnG2}, we plot the magnitude distributions for the G2-formulation together with the HR-formulation and disc-GAM. We observe how the signal constellation magnitudes have nearly the same distribution as the HR-formulation (with some oscillations) for low SNRs, but approaches the distribution of the disc-GAM for high SNRs. Thus, in the low SNR-region, the constellation approximate a complex Gaussian r.v., whereas in the entropy-limited region, a disc with a uniform distribution is approximated. In Fig.~\subref{fig:MagnG2-b}, we observe that the polynomial spiral power function has its limitation, and its square root approximate the function $\sqrt{\ln(N/(N-n))}$ with some oscillatory behavior. This suggests a non-polynomial approximation for the spiral power function, e.g. a one-parameter function that can be smoothly adapted between a bell-shape and a disc-shape.

The corresponding signal constellations for the G2-formulation optimized GB-GAM are also plotted in Fig. \ref{fig:ConstG2}. It is noted that the optimal constellation approximate a complex Gaussian r.v. at low SNR, and a uniform disc r.v. at high SNR.

\subsection{Probabilistic Bell-GAM}
\label{sec:Sec5p5d4d1d2}

\begin{figure}[tp!]
 \centering
 \vspace{-.4 cm}
 \includegraphics[width=9cm]{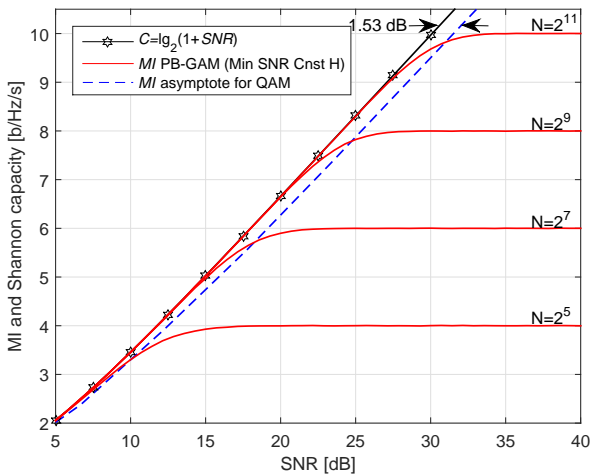}
  \caption{MI of PB-GAM (min SNR - entropy constrained) with $\log_2(N)=\{5,7,9,11\}$, QAM asymptote, and AWGN Shannon capacity.}
 \label{fig:Fig5p5dot5}
 \vspace{-0.0cm}
\end{figure}

\begin{figure}[tp!]
 \centering
 \vspace{-.4 cm}
 \includegraphics[width=9cm]{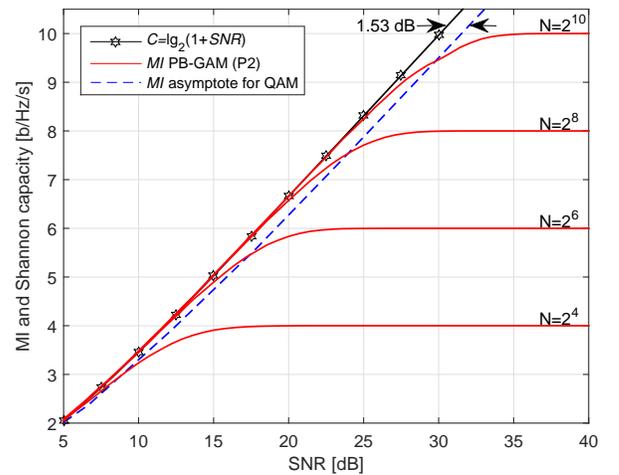}
  \caption{MI of PB-GAM (Formulation-P2) with $\log_2(N)=\{4,6,8,10\}$, QAM asymptote, and AWGN Shannon capacity.}
\label{fig:P2}
 \vspace{-0.0cm}
\end{figure}

Similar conclusion, as for the HR GB-GAM, Fig. \ref{fig:Fig5p5dot4}, is drawn for PB-GAM (min SNR - entropy constrained) shown in Fig. \ref{fig:Fig5p5dot5}. However, the MI for PB-GAM schemes is substantially higher, almost with a near ideal shape. But keep in mind, the scheme requires a larger $N$ for the same entropy as PG-GAM (high-rate), due to non-identical $p_n$. For the optimization formulation-P2, we show the MI-performance in Fig. \ref{fig:P2}. Here, $H=\log_2(N)$, and the probabilistic-shaping is optimized wrt the SNR. This is slightly better than than HR GB-GAM, but requires a shaper which adapt $p_n$ to the SNR.

\subsection{Overall Comparison}
\label{sec:Sec5p5d4d13}

\begin{figure}[tp!]
 \centering
 \vspace{-.4 cm}
 \includegraphics[width=9cm]{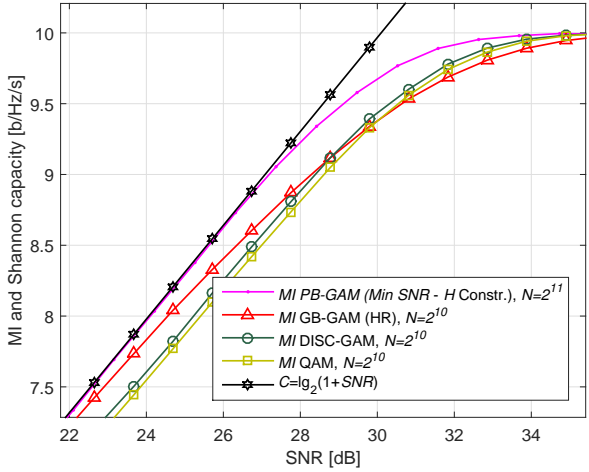}
 \caption{MI of disc-GAM, GB-GAM (HR), PB-GAM (min SNR - entropy constrained), QAM, and AWGN Shannon capacity.}
 \label{fig:Fig5p5dot6}
 \vspace{0.0cm}
\end{figure}

In Fig. \ref{fig:Fig5p5dot6}, we compare four different modulation schemes, disc-GAM, GB-GAM (high-rate), PB-GAM (min SNR - entropy constrained), and QAM, together with the AWGN Shannon capacity for high $N$. We note the SNR-gap of QAM and disc-GAM to the AWGN Shannon-capacity, and the $\approx0.2$ dB SNR gap between disc-GAM and QAM. We observe that when the MI approaches the entropy, disc-GAM, and QAM, perform better than GB-GAM (high-rate) scheme. Thus, GB-GAM (high-rate), while essentially optimal at lower MI, is suboptimal for $MI\approx H$.

\begin{figure}[tp!]
 \centering
 \vspace{-.4 cm}
 \includegraphics[width=9cm]{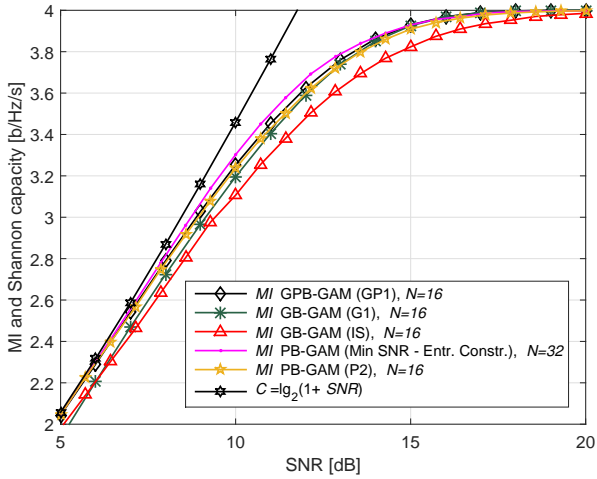}
 \caption{MI of GB-, PB-, and GPB-GAM variants, $N=16$.}
 \label{fig:Fig5p5dot6b}
 \vspace{0.0cm}
\end{figure}

In Fig. \ref{fig:Fig5p5dot6b},  the focus is shifted toward the MI-optimized scheme, but includes non-optimized schemes for reference. High-rate GB-GAM, which is not based on any optimization, performs the worst. GB-GAM with optimization formulation-G1 performs significantly better than GB-GAM (HR) when the MI is of the same order as the entropy, but is still the second worst. PB-GAM optimization formulation-P2, tuning the probabilities, comes next, and, overall, performs better than geometric-GAM in general. We also investigated PB-GAM optimization formulation P1, and it matches the MI-performance for P2. Hence, not shown here. When geometric and probabilistic shaping is combined, and optimized, as in GPB-GAM formulation GP1, it performs (as expected) better than either of GB-, and PB-GAM. The MI PB-GAM (min SNR - entropy constraint) performs the best here, at the cost of twice as many constellation points as the other schemes.

\subsection{Symbol Error Rate}
\label{SER}
\begin{figure}[tp!]
 \centering
 \vspace{-.4 cm}
 \includegraphics[width=9cm]{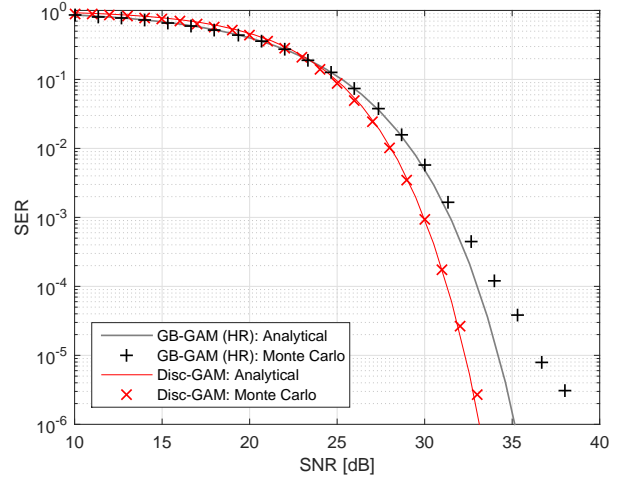}
  \caption{Analytical and Monte Carlo simulated SER for disc-GAM and GB-GAM (HR) with $N=256$.}
 \label{fig:SER}
 \vspace{-0.0cm}
\end{figure}

In Fig.  \ref{fig:SER}, we illustrate SER vs. SNR based on the analytical formulas for disc-GAM and GB-GAM (HR), alongside with Monte Carlo simulated SER. For disc-GAM, the analytical approximation and the simulated result matches relatively well. In the low-SNR region, the analytical model overestimate the SER somewhat. For GB-GAM (HR), we observe a very good match between analytical approximation and simulated result in the low-SNR region. In the high-SNR region, the analytical result deviates significantly from the simulated result, suggesting that the model can be improved.

\section{Summary and Conclusions} 
\label{Summary}
In this work, we have extended our newly proposed modulation  format (Golden angle modulation) that, with geometric or probabilistic-shaping, can approximate a circular-symmetric pdf of choice. We illustrated such approximations for the pdf of a continuous complex Gaussian r.v. For those schemes, the MI-performance asymptotically approaches the AWGN Shannon capacity as the number of signal constellation points grows, provided the MI is less than the entropy of the modulation scheme. This also implies that the shaping-loss vanishes, (0 dB). We gave several optimization frameworks, maximizing the MI under an average SNR constraint, and optionally also a PAPR constraint. The best MI-performance was achieved when both constellation point-geometry and -probabilities where optimized, and the second best appears to be when only the constellation point probabilities where optimized. A version of disc-GAM, with lower PAPR, was introduced, and which can serve as an alternative to PSK.

We believe that GAM may find many applications in transmitter resource limited links, such as space probe-to-earth, satellite, or mobile-to-basestation communication. This is so since high-data rate is desirable from the power-, energy-, and complexity-limited transmitter side, but higher complexity decoding is acceptable at the receiver side. Nevertheless, the performance of GAM is believed to be of interest for any communication system, wireless, optical, or wired. Numerous interesting extensions and applications of GAM have already been envisioned, studied, and will be presented within short.

\section{Appendix}

\subsection{Proof in Definition \ref{def:Def5p5d2}}
\label{app:App5p5dA1}
\begin{IEEEproof}
The constant $c_\textrm{disc}$ is determined from the average power constraint.
\begin{align}
\bar P&= \sum_{n=N_\textrm{l}}^{N_\textrm{h}} p_n
 r_n^2
 = c_\textrm{disc}^2 \sum_{n=N_\textrm{l}}^{N_\textrm{h}} \frac{n}{N}\notag\\
&=\frac{c_\textrm{disc}^2}{N} \left(\frac{N_\textrm{h}(N_\textrm{h}+1)}{2}-\frac{N_\textrm{h}(N_\textrm{h}-1)}{2}\right)\notag\\
 \Rightarrow \,  c_\textrm{disc}&=\sqrt{\frac{2\bar PN}{N_\textrm{h}(N_\textrm{h}-1)-N_\textrm{l}(N_\textrm{l}-1)}}.\notag
\end{align}
\end{IEEEproof}

\subsection{Proof in Theorem  \ref{thm:Thm5p5d1}}
\label{app:App5p5dA2d1}
\begin{IEEEproof}
The distribution in phase, for the constellation points, is already given by the $\mathbf{e}^{2\pi \varphi n}$-factor. However, the radial distribution need to be determined. A complex Gaussian r.v. with variance $\sigma^2$ is initially considered. The inverse sampling method, assumes a uniform continuous pdf on $(0,1)$. We modify the inverse sampling method and use a discrete uniform pmf at steps $n/N, \, n\in\{0, 1,\ldots,N-1\}$, where $N$ is assumed large (for the high-rate approximation).
\begin{align}
F(r_{n})
&=\int_0^{r_{n}}f_R(r)dr=\int_0^{r_{n}}\frac{\mathrm{e}^{-\frac{r^2}{\sigma^2}}}{\pi\sigma^2}2 \pi r \, \mathrm{d}r\notag\\
&=1- \mathrm{e}^{-\frac{r_n^2}{\sigma^2}}\notag.
\end{align}
Setting $F(r_n)=n/N$, and solving for $r_n$, yields
\begin{align}
r_n=\sigma\sqrt{\ln\left(\frac{N}{N-n}\right)}\notag .
\end{align}
Thus, the general solution for the signal constellation has the form $r_n=c_\textrm{gb} \sqrt{\ln{\left(\frac{N}{N-n}\right)}}$. The constant $c_\textrm{gb}$ is given by the average power constraint as follows,
\begin{align}
\bar P&=\sum_{n=0}^{N-1} p_n r_n^2=\frac{c_\textrm{gb}^2}{N}\sum_{n=0}^{N-1}\ln{\left(\frac{N}{N-n}\right)},\notag\\
\Rightarrow c_\textrm{gb}&=\sqrt{\frac{N \bar P}{\sum_{n=0}^{N-1}\ln{\left(\frac{N}{N-n}\right)}}}
=\sqrt{\frac{N \bar P}{N\ln{N}-\ln{N!}}}\notag.
\end{align}
\end{IEEEproof}

\subsection{Discussion on MI-optimization Formulations}
\label{app:App5p5dA2}

\subsubsection{Optimization Formulation-G1}
\label{app:App5p5dA2d1}

It is instructive to see why formulation-G1 is hard to solve. Therefore, we derive the optimization criteria, while  omitting the troublesome magnitude ordering condition $r_{n+1}\geq r_n$, below. The MI, $I(Y;X)$, for a general GAM-constellation is already given in Section \ref{sec:Sec5p5d2d3d1}. The Lagrangian, omitting the condition $r_{n+1}\geq r_n$, is equivalent to maximize the differential entropy of $Y$ given the SNR condition, i.e.
\begin{align}
\Lambda= h(Y)-\lambda\left(\frac{1}{N\sigma^2}\sum_{n=0}^{N-1}r_n^2-S\right),
\end{align}
which yields the optimality conditions
\begin{align}
\frac{\mathrm{d} h(Y)}{\mathrm{d} r_n}-\frac{2\lambda}{N\sigma^2}r_n=0, \forall n,
\end{align}
where the derivative of the differential entropy is
\begin{align}
\frac{\mathrm{d} h(Y)}{\mathrm{d} r_n}
&=-\frac{2}{\ln(2)N\pi\sigma^2}\int_{\mathbb{C}} \mathrm{e}^{-\frac{|y-x_n|^2}{\sigma^2}} \left(1+\ln(f_Y)\right) \notag\\
\times& \left(\mathrm{Re}\{y\mathrm{e}^{-i2\pi \varphi n}\}-r_n\right) \, \mathrm{d}y.
\end{align}
Hence, the $N$ Lagrangian optimality conditions reduces to
\begin{align}
&\int_{\mathbb{C}}  \ln(\mathrm{e}f_Y)  \left(\mathrm{Re}\{y\mathrm{e}^{-i2\pi \varphi n}\}-r_n\right)\notag\\
\times &\mathrm{e}^{-\frac{|y-r_n\mathrm{e}^{i2\pi \varphi n}|^2}{\sigma^2}} \, \mathrm{d}y+\tilde \lambda r_n=0,
\end{align}
where $\tilde \lambda= \pi\ln{(2)} \lambda  $. In addition, the ordering condition $r_{n+1}\geq r_n$, and the SNR condition, must be considered. Unfortunately, we can not solve this non-linear system of equations analytically.

\subsubsection{Optimization Formulation-G2}
\label{app:App5p5dA2d2}
As in formulation-G1, the derivative of the MI wrt $c_k$  can be replaced with derivative of the differential entropy $h(Y)$ wrt $c_k$. We have
\begin{align}
\frac{\mathrm{d} h(Y)}{\mathrm{d} c_k}
&=-\frac{1}{\ln(2)}\int_{\mathbb{C}} \ln(\mathrm{e}f_Y)\sum_{n=0}^{N-1} \frac{1}{N\pi\sigma^2}\mathrm{e}^{-\frac{|y-\sqrt{f_\textrm{P}(n/N)} \mathrm{e}^{i2\pi \varphi n}|^2}{\sigma^2}}\notag\\
&\times \frac{1}{\sigma^2}\left(\frac{\mathrm{Re}\{y\mathrm{e}^{-i2\pi \varphi n}\}}{\sqrt{f_\textrm{P}(n/N)}}-1\right) \frac{\mathrm{d} f_\textrm{P}(n/N)}{\mathrm{d} c_k} \, \mathrm{d}y .
\end{align}
The derivative of the Lagrangian wrt $c_k$ yields
\begin{align}
&-\int_{\mathbb{C}} \frac{\ln(\mathrm{e}f_Y)}{\ln(2)N\pi\sigma^4}\sum_{n=0}^{N-1} \mathrm{e}^{-\frac{|y-\sqrt{f_\textrm{P}(n/N)} \mathrm{e}^{i2\pi \varphi n}|^2}{\sigma^2}}\notag\\
\times& \left(\frac{\mathrm{Re}\{y\mathrm{e}^{-i2\pi \varphi n}\}}{\sqrt{f_\textrm{P}(n/N)}}-1\right)\left(\frac{n}{N}\right)^k \, \mathrm{d}y=\lambda d_k.
\end{align}
Unfortunately, solving for $c_k$ is analytically untractable.

\subsection{Proof of Theorem \ref{thm:Thm5p5d2}}
\label{app:App5p5dA3}
\begin{IEEEproof}
The Lagrange function to optimize wrt $p_n$ is
\begin{align}
\Lambda&=\sum_{n=1}^{N}p_n n+\lambda_1\left(1-\sum_{n=1}^{N}p_n\right)\notag\\
&+
\lambda_2\left(H_\textrm{pbse}+\sum_{n=1}^{N}p_n \ln(p_n)\right)\notag
\end{align}
where $\lambda_1$ and $\lambda_2$ are Lagrange parameters to be determined.

Taking the derivative of the Lagrange function, equating to zero, $\frac{\mathrm{d} \Lambda}{\mathrm{d} p_n}=0$, yields the optimality condition
\begin{align}
&n-\lambda_1+\lambda_2(\ln(p_n)+1)=0,\notag\\
&\Rightarrow p_n=\mathrm{e}^{-\frac{n-\lambda_1}{\lambda_2}-1}\Rightarrow p_n=c_\textrm{p} \xi^n \notag
\end{align}
where $c_\textrm{p}$ and $\xi$ are parameters, related to $\lambda_1$ and $\lambda_2$, to be determined.

The constant $c_\textrm{p}$ is found from the unit-sum probability condition.
\begin{align}
1&=\sum_{n=1}^{N}p_n
=\sum_{n=1}^{N}c_\textrm{p} \xi^n
=c_\textrm{p} \xi\frac{1-\xi^N}{1-\xi}\notag\\
\Rightarrow c_\textrm{p}&=\frac{1}{\xi}\frac{1-\xi}{1-\xi^N} \Rightarrow p_n=\frac{1}{\xi}\frac{1-\xi}{1-\xi^N} \xi^n.\notag
\end{align}

The constant $c_\textrm{pbse}$ is found by using the average SNR $S$.
\begin{align}
S\sigma^2&=\sum_{n=1}^{N}p_n r_n^2
=c_\textrm{pbse}^2\sum_{n=1}^{N}\frac{1}{\xi} \frac{1-\xi}{1-\xi^N} \xi^n n\notag\\
&=c_\textrm{pbse}^2\left(\frac{1}{1-\xi} -\frac{N\xi^N}{1-\xi^N}\right)\notag\\
\Rightarrow c_\textrm{pbse}
&=\sqrt{S \sigma^2 \left(\frac{1}{1-\xi}-\frac{N\xi^N}{1-\xi^N}\right)^{-1}}\notag
\end{align}
In the above, we can replace $S\sigma^2$ with $\bar P$.
Using $p_n$, the entropy is
\begin{align}
H_\textrm{pbse}
&=-\sum_{n=1}^N p_n \ln(p_n)\notag\\
&=-\sum_{n=1}^N \frac{1}{\xi}\frac{1-\xi}{1-\xi^N} \xi^n \ln \left( \frac{1}{\xi}\frac{1-\xi}{1-\xi^N} \xi^n\right)\notag\\
&=-\sum_{n=1}^N \frac{1}{\xi}\frac{1-\xi}{1-\xi^N} \xi^n \ln \left(\frac{1}{\xi} \frac{1-\xi}{1-\xi^N}\right)\notag\\
&-\sum_{n=1}^N \frac{1}{\xi}\frac{1-\xi}{1-\xi^N} \xi^n \ln \left( \xi^n\right)\notag\\
&=-\ln \left(\frac{1}{\xi} \frac{1-\xi}{1-\xi^N}\right)-\frac{\ln (\xi)}{\xi}\frac{1-\xi}{1-\xi^N}\sum_{n=1}^{N} n\xi^n \notag\\
&=-\ln \left(\frac{1}{\xi}  \frac{1-\xi}{1-\xi^N}\right)-\left(\frac{1}{1-\xi} -\frac{N\xi^N}{1-\xi^N}\right)\ln (\xi)\notag\\
&=-\ln \left(  \frac{1-\xi}{1-\xi^N}\right)+\left(\frac{N\xi^N}{1-\xi^N}-\frac{\xi}{1-\xi}\right)\ln (\xi).\notag
\end{align}

\end{IEEEproof}


\end{document}